\documentstyle[12pt,epsfig]{article}
\input epsf
\begin{document}
\title{EFFECTIVE LAGRANGIANS INDUCED BY THE
ANOMALOUS WESS-ZUMINO ACTION AND $I^G(J^{PC})=1^-(1^{-+})$ EXOTIC
STATES \thanks{Parallel session talk given at HADRON 2001, Protvino, Russia,
August 28, 2001.}}
\author{N.N. Achasov and G.N. Shestakov\\
{\it Laboratory of Theoretical Physics, S.L. Sobolev Institute}\\
{\it for Mathematics, 630090, Novosibirsk, Russia}}
\date{}
\maketitle\begin{abstract}
A simple dynamical model for the exotic waves with
$I^G(J^{PC})=1^-(1^{-+})$ in the reactions $\rho\pi\to\rho\pi$,
$\rho\pi\to\eta\pi$, $\rho\pi\to\eta'\pi$, $ \rho\pi\to(K^*\bar
K+\bar K^*K)$, and in the related ones, is constructed beyond the
scope of the quark-gluon approach. The model satisfies unitarity
and analyticity and uses as a ``priming"\ the anomalous
non-diagonal $VPPP$ interaction which couples together the four
channels $\rho\pi$, $\eta\pi$, $ \eta'\pi$, and $K^*\bar K+\bar
K^*K$. The possibility of the resonance-like behavior of the
$I^G(J^{PC})=1^-(1^{-+})$ amplitudes belonging to the $\{10\}-
\{\bar{10}\}$ and $\{8\}$ representations of $SU(3)$ as well as
their mixing is demonstrated explicitly in the 1.3--1.6 GeV mass
range which, according to the current experiments, is really rich
in exotics.
\end{abstract}

\begin{center}{\bf INTRODUCTION}\end{center} 

Phantoms of manifestly exotic $\pi_1$ states with
$I^G(J^{PC})=1^-(1^{-+})$ have more and more agitated the
experimental and theoretical communities [1-3]. They were
discovered in the 1.3--1.6 GeV mass range in the $\eta\pi$,
$\eta'\pi$, $ \rho\pi$, $b_1\pi$, and $f_1\pi$ systems produced in
$\pi^-p$ collisions at high energies and in $N\bar N$ annihilation
at rest in the GAMS, KEK, VES, CB, and BNL experiments [1-3].

The first evidence for the possible existence of an exotic
$1^{-+}$ state coupled to the $\eta\pi$ and $\rho\pi$ channels and
belonging to the icosuplet representation of $SU(3)$ was obtained
by J. Schechter and S. Okubo about 37 years ago with the bootstrap
technique [4].

Recently theoretical considerations concerning the mass spectra
and decay properties of exotic hadrons have been based, in the
main, on the MIT-bag model, constituent gluon model, flux-tube
model, QCD sum rules, lattice calculations, and various selection
rules.

Current algebra and effective Lagrangians are also important
sources of theoretical information on exotic partial waves. It is
sufficient to remember the prediction obtained within the
framework of these approaches for the $\pi\pi$ $S$-wave scattering
length with isospin $I=2$. There also exist a good many of the
model constructions which show that the low-energy contributions
calculated within the effective chiral Lagrangians framework may
in principle transform with increasing energy into resonances with
the experimentally established parameters. The important
ingredient of all these models is the successfully selected
unitarization scheme for the original chiral amplitudes which is
used to match the low-energy and resonance regions. Such models
are well known, for example, for the $\pi\pi$ scattering channels
involving the $\sigma$ and $\rho$ resonances (see, for example,
Ref. [5]). In the present work we continue in this way and
construct

\begin{center}{\bf A MODEL FOR THE \mbox{\boldmath$I^G(J^{PC})=1^-(1^{-+})$}
WAVES IN THE REACTIONS \mbox{\boldmath$VP\to PP$}, \mbox{\boldmath$\ PP\to PP
$}, \ AND \mbox{\boldmath$\ VP\to VP$}}\end{center} using, as the starting
point, the following anomalous effective Lagrangian for point-like
$VPPP$ interaction of the vector $(V)$ and pseudoscalar $(P)$
mesons
$$L(VPPP)=ih\,\epsilon_{\mu\nu\tau\kappa}\mbox{Tr}(\hat V^\mu\partial^\nu\hat P
\partial^\tau\hat P\partial^\kappa\hat P)+i\sqrt{1/3}\,h'\,\epsilon_{\mu\nu\tau
\kappa}\mbox{Tr}(\hat V^\mu\partial^\nu\hat P\partial^\tau\hat
P)\partial^ \kappa\eta_0\,,$$ where $h$ and $h'$ are the coupling
constants, $\hat P=\sum _{a=1}^{8}\lambda_aP_a/\sqrt{2}$, $\ \hat
V^\mu=\sum_{a=0}^{8}\lambda_aV^\mu_a/ \sqrt{2}$, and $\lambda_a$
are the Gell-Mann matrices. This Lagrangian induced by the
anomalous Wess-Zumino action and generates the tree exotic
amplitudes with $I^G(J^{PC})=1^-(1^{-+})$ for the inelastic
reactions $\rho\pi\to\eta_8\pi,\ K^*\bar K\to\eta_8\pi,\ \bar
K^*K\to\eta _8\pi$ belonging to the $\{10\}-\{\bar{10}\}$
representation of $SU(3)$ (in this case, there are, at least, the
$qq\bar q\bar q$ states in the $s$-channel) and the tree
amplitudes for the reactions $\rho\pi\to\eta_0\pi,\ K^*\bar
K\to\eta_0\pi,\ \bar K^*K\to\eta_0\pi$ belonging to the $\{8\}$
representation of $SU(3)$ (in this case, there are both $qq\bar
q\bar q$ and $q \bar qg$ states in the $s$-channel).\,\footnote{We
use the pseudoscalar octet-singlet ($\eta_8$-$\eta_0$) mixing
angle $\theta_P\approx-20^\circ$.} In the next orders, these tree
amplitudes induce as well the exotic ones for the elastic
processes $\rho\pi\to\rho\pi$, \ $\eta\pi\to\eta\pi$, and so on.
In this connection it is of interest to consider the following
$4\times4$ system of scattering amplitudes for the coupled exotic
channels of the reactions $VP\to VP,\ VP \leftrightarrow PP$ and
$PP\to PP$:
$$T_{ij}=\left[\begin{array}{llll}
T(\rho\pi\to\rho\pi) & T(\rho\pi\to\eta\pi) &
T(\rho\pi\to\eta'\pi) &
T(\rho\pi\to K^*K)\\
T(\eta\pi\to\rho\pi) & T(\eta\pi\to\eta\pi) &
T(\eta\pi\to\eta'\pi) &
T(\eta\pi\to K^*K)\\
T(\eta'\pi\to\rho\pi) & T(\eta'\pi\to\eta\pi) &
T(\eta'\pi\to\eta'\pi) &
T(\eta'\pi\to K^*K)\\
T(K^*K\to\rho\pi) & T(K^*K\to\eta\pi) & T(K^*K\to\eta'\pi) & T(K^*K\to K^*K)\\
\end{array}\right].$$
The subscripts $i,j=1,2,3,4$ are the labels of the $\rho\pi,\
\eta\pi, \ \eta'\pi$, and $K^*K$ channels, respectively (the
abbreviation $K^*K$ implies just the $\bar K^*K$ and $K^*\bar K$
channels).

We consider three natural limiting (in the sense of $SU(3)$
symmetry) cases: (i) $h'=0$, i.e., when all exotic amplitudes
belong to the $\{10\}-\{\bar{10}\}$ representation of $SU(3)$;
(ii) $h=0$, i.e., when all exotic amplitudes belong to the octet
representation of $SU(3)$; and (iii) $h'=h$, when the original
$VPPP$ interaction possesses nonet symmetry with respect to the
$0^-$ mesons.

To obtain the unitarized amplitudes in coupled channels, we sum up
all the possible chains of the $s$-channel loop diagrams the
typical examples of which are given below.

\setlength{\unitlength}{1mm}\thicklines
\begin{picture}(140,32)(-4,63.5)
\put(1,86){\line(1,-1){16}} \put(1,70){\line(1,1){16}}
\put(20,77){+} \put(73,77){+} \put(126,77){+\ \,.\,.\,.}
\put(0,88){$\rho^0$}\put(17,88){$\eta$}
\put(25,88){$\rho^0$}\put(40,88){$\eta$}\put(54,88){$\rho$}\put(70,88){$\eta$}
\put(78,88){$\rho^0$}\put(93,88){$\eta$}\put(107,88){$K^*$}\put(123,88){$\eta$}
\put(0,66){$\pi^-$}\put(17,66){$\pi^-$}
\put(25,66){$\pi^-$}\put(40,66){$\pi^-$}\put(54,66){$\pi$}\put(70,66){$\pi^-$}
\put(78,66){$\pi^-$}\put(93,66){$\pi^-$}\put(107,66){$K$}\put(123,66){$\pi^-$}
\put(26,86){\line(1,-1){8}} \put(26,70){\line(1,1){8}}
\put(41,78){\circle{20}} \put(55,78){\circle{20}}
\put(62,78){\line(1,-1){8}} \put(62,78){\line(1,1){8}}
\put(79,86){\line(1,-1){8}} \put(79,70){\line(1,1){8}}
\put(94,78){\circle{20}} \put(108,78){\circle{20}}
\put(115,78){\line(1,-1){8}} \put(115,78){\line(1,1){8}}
\end{picture}\\
\begin{figure}\centerline{
\includegraphics[height=.75\textheight]{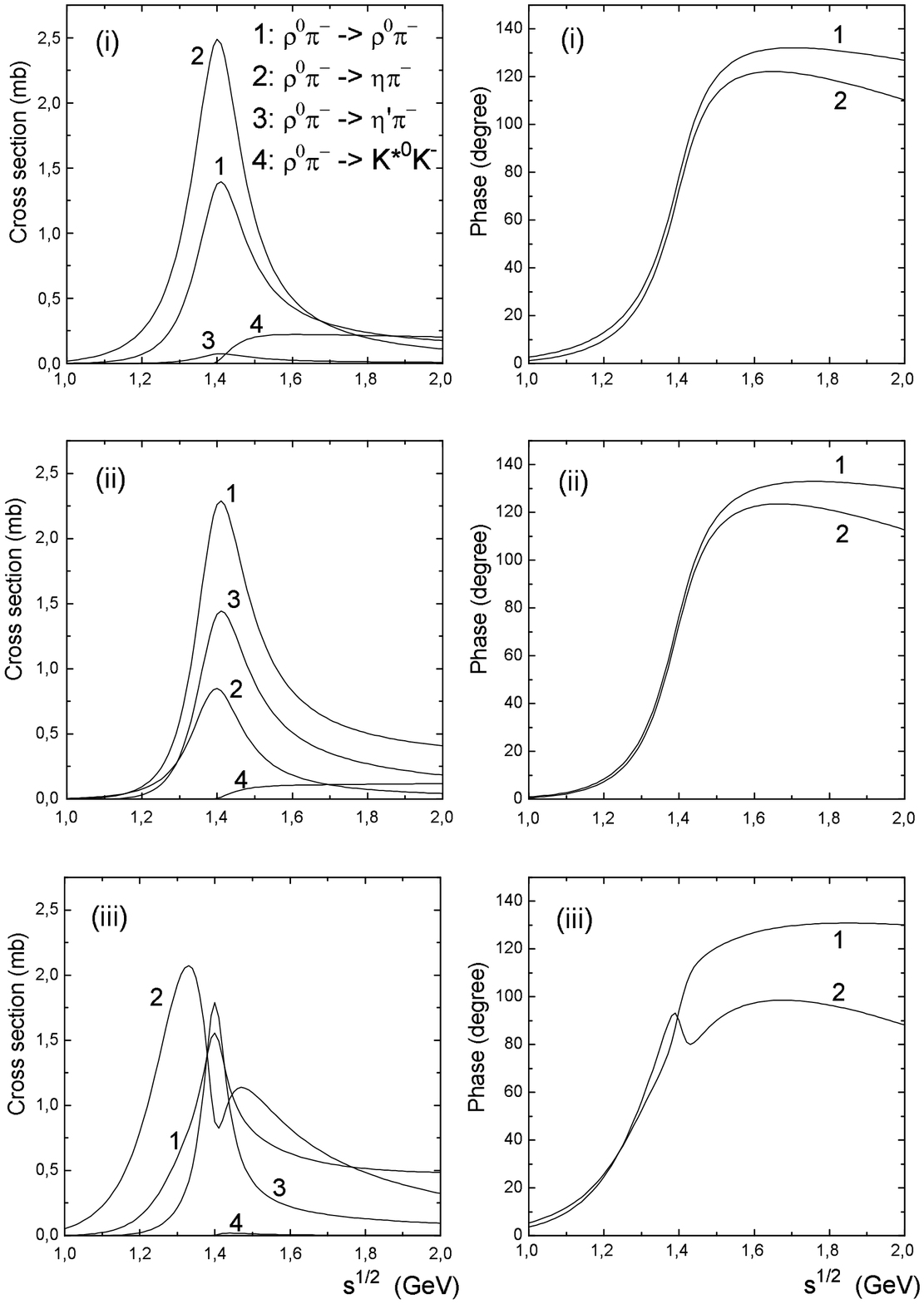}}
\caption{\ The cross sections of the reactions
$\rho^0\pi^-\to\rho^0\pi^-$, $\rho^0\pi^-\to\eta\pi^-$,
$\rho^0\pi^-\to\eta'\pi^-$, and $\rho^0\pi^-\to K^{* 0}K^-$ and
the phases of the $\rho\pi\to\rho\pi$ and $\rho\pi\to\eta\pi$
amplitudes. The correspondence between the curve numbers and the
reaction channels is shown just in the figure. $\tilde h=0.10746$,
$\tilde h'=0$, $C_{11}=0.17$ GeV$^2 $, $C_{12}=1.25$ GeV$^2$ in
case (i), $\tilde h'=0.10746$, $\tilde h=0$, $C_{11}=0.34$ GeV$
^2$, $C_{12}=0.67$ GeV$^2$ in case (ii), and $\tilde h=\tilde
h'=0.10746$, $C _{11}=0.49$ GeV$^2$, $C_{12}=0.5$ GeV$^2$ in case
(iii).}\end{figure}
It is the well known field theory way of the unitarization. The
relevant summation can be easily carried out by using the matrix
equation $\tilde T_{ij}=h_{ij}+h_{im}\Pi_{mn}\tilde T_{nj}$ for
the auxiliary invariant amplitudes $\tilde T_{ij}$, the solution
of which has the form: $\tilde T_{ij}=[\,(\hat1-\hat
h\hat\Pi)^{-1}\,]_{im}\,h_{mj}.$
Here $ h_{ij}$ is the matrix of the coupling constants generated
by the Lagrangian and $\Pi_{ij}=\delta_{ij}\Pi_j$ is the diagonal
matrix of the loops, $\Pi_{i}=sF_i/(6\pi)$ for the $VP$ loops
(i=1,4) and $\Pi_{i}=F_i/(24\pi)$ for the $PP$ loops (i=2,3), the
functions $F_i$ are defined by the doubly subtracted dispersion
integrals $$F_i=C_{1i}+sC_{2i}+
\frac{s^2}{\pi}\int\limits_{m^2_{i+}}^{\infty}\frac{[P_i(s')]^3\
ds'}{\sqrt{s'} \,s'^{\,2}(s'-s-i\varepsilon)}\,,$$ where
$\sqrt{s}$ is the invariant mass of two-body systems, $P_i(s')$
and $m_{i+}$ are the particle momentum and the sum of the 
particle masses in the $i$ intermediate state.

Next, we define $D=\mbox{det}(\hat1-\hat h\hat\Pi)$. It is clear
that all the physical amplitudes $T_{ij}$ are proportional to
$1/D$. Thus a simplest way to discover ``by hand"\ a possible
resonance situation is that to find zero of Re$(D)$ at fixed
values of $h$, $h'$ and $\sqrt{s}$ (for example, at
$\sqrt{s}=1.43$ GeV). Leaving the potentialities of the model
almost unchanged, we assume that $C_{11}=C_{14}$ and
$C_{21}=C_{24}=0$ for the $VP$ loops, $C_{12}=C_{13}$ and
$C_{22}=C_{23}=0$ for the $PP$ loops. Thus, in most considered
variants, we used as the essential free parameters only two
subtraction constants $C_{11}$ and $C_{12}$. As for the coupling
constant $h$, one may claim [6] that it is not too large in the
scale defined by the combination
$2g_{\rho\pi\pi}g_{\omega\rho\pi}/m^2_\rho \approx284$
GeV$^{-3}\,$, namely, that $\vert\tilde h=F^3_\pi h\vert\leq0.4$
(where $F _\pi\approx130$ MeV), and we are guided by the values of
$\tilde h$ (and $ \tilde h'=F^3_\pi h')$ near 0.1.

Figure 1 shows the typical energy dependences, which occur in our
model for cases (i), (ii), and (iii), for the four reaction cross
sections $\sigma(\rho^0\pi^-\to\rho^0\pi^-) $,
$\sigma(\rho^0\pi^-\to\eta\pi^-)$,
$\sigma(\rho^0\pi^-\to\eta'\pi^-)$, and $ \sigma(\rho^0\pi^-\to
K^{*0}K^-)$ and for the phases of the $\rho\pi\to\rho\pi$ and
$\rho\pi\to\eta\pi$ amplitudes. They clearly demonstrate the
resonance effects found in the invariant mass region 1.3--1.4 GeV
(a similar resonance picture is also obtained for the 1.5--1.6 GeV
mass region [7]). Furthermore, the comparison of the obtained
cross section values (see Fig. 1) with those of the conventional
$a_2(1320)$ resonance production, $\sigma(\rho^0 \pi^-\to
a_2\to\rho^0\pi^-)\approx5.7$ mb and $\sigma(\rho^0\pi^-\to
a_2\to\eta \pi^-)\approx2.36$ mb at $\sqrt{s}=m_{a_2}=1.32$ GeV,
indicates conclusively that we are certainly dealing with the
resonance-like behavior of the $I^G(J^{PC })=1^-(1^{-+})$ exotic
waves, at least, in the $\rho\pi,\ \eta\pi$, and $\eta' \pi$
channels. Summarizing we conclude that our calculation (see Ref.
[7] for details) gives a further new reason in favor of the
plausibility of the existence of an explicitly exotic $\pi_1$
resonance in the mass range 1.3--1.6 GeV.

\end{document}